\documentclass[12pt,oneside,final,a4paper]{article}

\newcommand{\epigraph}[3]{\par
\hfill\parbox{0.6\textwidth}{\footnotesize #1 \par \hfil #2 
\textit{#3}}\par}%

\title{V\"axj\"o Interpretation of Quantum Mechanics}

\author{Andrei Khrennikov\\
International Center for Mathematical Modeling\\ in Physics and Cognitive sciences \\
MSI, V\"axj\"o
University, S-35195, Sweden}

\begin{document}
\maketitle

\begin{abstract}
We present critical arguments against individual interpretation of Bohr's
complementarity and Heisenberg's uncertainty principles.
Statistical interpretation of these principles is discussed in the 
contextual framework. We support the possibility to 
use Statistical Contextual Realist Interpretation of quantum 
formalism.  In spite of all {\bf no-go} theorems (e.g., von Neumann, Kochen and Specker,...,
Bell,...), recently (quant-ph/0306003 and 0306069)
we constructed a realist basis of quantum mechanics. In our model both
classical and quantum spaces are rough images of the fundamental
{\bf prespace.} Quantum mechanics cannot be reduced to classical one. Both classical and quantum
representations induce reductions of prespace information.
\end{abstract}

\epigraph{"Often we enter the unknown edifice of a new scientific discipline through a lesser gate that 
leads us into a side passage. It may take us a long while to find our way to the main portal and to view 
the whole structure in its proper perspective".}{ E. Schr\"odinger}{}

\section{Introduction}

In this paper we present some critical arguments against individual understanding 
of Bohr's complementarity and Heisenberg's uncertainty principles.
It is argued the only possible interpretation of these principles is the 
statistical contextual one. In particular, in the opposite to orthodox Copenhagen
views such interpretation does not imply the impossibility to assign to elementary
particles\footnote{The reader may protest against the use 
of the notion `particle'. It might be better to use e.g. `quantum system.' But I dislike to use `quantum'
regarding to real physical objects. `Quantum' has become merely the symbol of  a rather
special mathematical formalism.}  objective properties; including position and momentum. Of course, such 
ontic description of physical reality might not be provided by quantum formalism. In particular,
it is clear that quantum formalism does not provide the complete description of 
physical reality. 

We argue that the most natural interpretation of physical theories (in particular,
quantum formalism) is Statistical Contextual Realist Interpretation.

\section{Contextual statistical realist interpretation of physical theories}

Quantum theory as well as classical physics is used to describe 
properties of pairs
\begin{equation}
\label{H1}
\pi=(\mbox{physical system}, \mbox{measuring device}).
\end{equation}
I do not think that understanding of this fact, {\it contextual structure of physical theories},
really  was Bohr's invention. It was clear to everybody that physical observables 
are related to properties of physical systems as well as measuring devices. 
The main invention of N. Bohr was not contextuality, but {\it complementarity.} Bohr's greatest
contribution was the recognition of the fact that there exist complementary experimental arrangements and
hence complementary, incompatible, pairs $\pi_1, \pi_2$ of form (\ref{H1}). I think nobody can be against the recognition
of such a possibility. Why not? Why must all contexts, complexes of physical conditions, be 
coexisting? Contextuality and complementarity are two well understandable principles (not only of quantum
physics, but physics in general). 

The real problem was that N. Bohr as well as W. Heisenberg 
(but merely further generations of their adherents) did not pay attention that quantum 
complementarity is the experimental
fact concerning pairs 
\begin{equation}
\label{H2}
\pi=(\mbox{elementary particle}, \mbox{macroscopic measuring device})
\end{equation}
and not elementary particles by their self. It is a pity that the greatest promoters of contextualism forgot
about contextual basis of complementarity. 

If we remember about contextuality discussing complementarity we see that complementarity of
contexts in quantum physics
does not imply complementarity of 
corresponding objective properties (of elementary particles) contributing into such observables.
In particular, contextual complementarity does not imply that elementary particles do not 
have objective properties at all. In particular, there are no reasons to suppose that it is impossible
to provide a kind of hidden variable, HV, description ({\it ontic description}, see e.g. [1]) for these 
objective properties. Mathematically the pair $\pi$ can be described by two variables $\lambda_s$ and
$\lambda_d$ representing objective states of a physical system and measuring device. So a physical
observable is represented as a function 
\begin{equation}
\label{H3}
A=A(\lambda_s;\lambda_d).
\end{equation}
In general, $\lambda_s$ can be represented as a vector with numerical coordinates. Numerical does not mean 
`real number'. It can be as well e.g. $p$-adic number, see [2]. So the possibility of HV-description
does not imply continuous infinitely divisible representation of physical reality. In principle,
classical physical description is often identified with the description based on the real-number
`continuous' model. We would not like to use such a restrictive, XVIIIth, XIXth century and the first half of 
XXth century, interpretation of classical physics. A classical physical model is every model 
that provides objective description of phenomena -- representation of physical observables in 
form (\ref{H3}). I think that mixing of `classical' and `continuous' was one of main roots
of vague interpreting of results of experiments with elementary particles. 
Thus the development of alternative (nonreal, noncontinuous) classical models, e.g. $p$-adic
[3], [4], might play important role in clarification of foundations of quantum theory.

On the other hand, an adherent of N. Bohr would argue that 
"Such a separation and, hence, the description of (properties of)
quantum objects and processes themselves (as opposed to certain effects of their interaction with measuring
instruments upon latter) are impossible in Bohr's interpretation," [5]. 

I think that the origin of such an interpretation of complementarity by N. Bohr
was the individual interpretation of Heisenberg's uncertainty principle:
\begin{equation}
\label{H4}
\Delta q \Delta p \geq \frac{h}{2}.
\end{equation}
Close relation between Bohr's complementarity principle and Heisenberg's uncertainty 
principle is well know.
For two years (1926-1927) N. Bohr could not present 
any model that could explain both corpuscular behaviour (black body radiation, photoelectric effect) 
and wavelike behaviour (two slit experiment, diffraction) of elementary particles.
You can read e.g. in the book [6] how heavy this thinking process was for N. Bohr - really a kind of mental disaster.
Only after the derivation by W. Heisenberg of the uncertainty principle,
N. Bohr proposed a new model of physical reality based on the principle of {\it complementarity.} 
Unfortunately, Heisenberg's uncertainty relation was interpreted as the relation for
an individual elementary particle. The main problem was mixing by W. Heisenberg
of {\it {individual}} and {\it {statistical}} 
uncertainty. For example, in his famous book [7] 
he discussed the uncertainty principle as a relation for an individual system, but derived this
principle by using statistical methods! 

The roots of such individual complementarity can be found already in the first work of W. Heisenberg [8].
At the beginning Heisenberg's quantization procedure was not  statistical one. It seems, see [8],
that W. Heisenberg was sure that he found equations for observables related to individual physical systems.
It was a rather common point view: in classical mechanics the position and momentum of an
individual (!) system are described by real numbers, in quantum mechanics -- by matrices. 
W. Heisenberg rightly underlined that matrice-description could not be used for describing objective
position and momentum of electron. There (as everywhere in XVIIIth -- XXth centuries physics) objective reality
was, in fact, identified with the continuous (real number) mathematical model of physical reality.
Impossibility to create a continuous real number model for motion of electron in Bohr's model of atom was considered
by Heisenberg (and many others) as impossibility to assign objective properties to electrons.
It was a rather strange passage. But we understand that it was the beginning of XXth century
and W. Heisenberg used the standard mathematical image of physical reality. However, we
would like to remark that even at that time there were attempts to modify continuous mathematical
model to reproduce quantum effects, see e.g. Bohr, Kramers, Slater [9] on classical-like 
quantization based on {\it difference equations} (instead of differential). In fact, this model 
stimulated M. Born to introduce the term quantum mechanics. In any case the absence of 
continuous classical model for motion of electron in Bohr's atom does not imply impossibility
to create other, noncontinuous, classical (causal deterministic) models. Moreover, considerations
of W. Heisenberg in [8] even did not imply impossibility to create continuous classical model --
as it was claimed by W. Heisenberg and then by N. Bohr. The story is much simpler: first Bohr
tried to create such a thing, but could not; then Heisenberg, with the same result. After this
it was claimed that such a model did not exist. And what is the most interesting: not only for
Bohr's model of atom (well it might be), but for any other model... I cannot understand this kind of 
`quantum logic'.  

Really W. Heisenberg proposed some mathematical model for some class of observations of position
and momentum of electron. These observations satisfy to the uncertainty relation. It is not clear
why we cannot present other mathematical models for some other observations of position and momentum
of electron that would violate this relation? Of course, if we relate Heisenberg's position 
and momentum to individual electron, then such individualization plays the role of objectification
of these quantities. It is rather strange logical circle, but it seems that it was done by 
W. Heisenberg and N. Bohr. Finally, this objectification in combination with the uncertainty
relation implies (for W. Heisenberg and N. Bohr) impossibility to consider
other position and momentum variables, distinct from Heisenberg's ones. On the other hand,
if we use statistical interpretation of uncertainty relation, then there are no reasons for
such NO-GO conclusions. Well, we could not prepare statistical ensembles with small
dispersions for two variables introduced by W. Heisenberg. But  we should have great imagination
to make Heisenberg-Bohr conclusions.

Unfortunately, many clear mind scientists used and still use Heisenberg-Bohr's `quantum logic'.
For instance, the fundamental paper of A. Zeilinger []
gives us an excelent example of the modern representation of this logic. 
In principle, A. Zeilinger is looking for a new quantum paradigm. He correctly 
underlines that the situation in quantum theory, especially large diversity of
interpretations, is not so natural. That in the opposite to e.g. theory of 
relativity, there is no quantum analogue of the {\it principle of relativity.} 
However, it seems that, for Zeilinger as well as for many other scientists looking
for reconsideration of quantum foundations, such a reconsideration could (and moreover
should) be performed as some addition to the orthodox Copenhagen.
A. Zeilinger started (as always in this story) with the correct statement that 
we are dealing with {\it a quantum phenomenon as the whole entity which comprises both the observed
quantum system and the classical measuring apparata.} No doubts! The formalism 
of quantum mechanics (statistical formalism) deals with such a phenomenon. 
However, then he continued: {\it It
is especially impossible in principle to predict with certainty both
through which slit an elementary particle will go and where it will appear
in the interference pattern.} Well, we still can interpret this statement 
in the correct way: quantum formalism does not give us such a possibility. 
Unfortunately, the latter understanding was impossible to orthdox Copenhagen,
since (by unclear reasons) it was supposed quantum theory provided the complete
description of physical reality.\footnote{I think that the paper of 
Einstein, Podolsky and Rosen, {\it Can quantum mechanical description
of physical reality be considered complete?}, [], was directed precisely against
Heisenberg-Bohr `quantum logic' that was, in fact,  based on the idea that 
quantum theory is complete.} So, Zeilinger continued: {\it I propose that 
this impossibility to describe
the random individual process within quantum mechanics in a complete way is
a fundamental limitation of the program of modern science to arrive at a
description of the world in every detail.} This is the great manifistation of Copenhagen
NO-GO.

Another important story that stimulated Bohr's complementarity thinking 
was Shr\"odinger's quantum story. It is important to recall that E. Shr\"odinger
was sure that he discovered totally new method of quantization [10],
"Quantization as the problem for eigenvalues." In his first paper  [10] he did not
refer to Heisenberg's paper [8]; in the second paper [11] he made a short reference
in the sense that W. Heisenberg proposed some other method of quantization
that was totally  different from Shr\"odinger's one. It is well known that 
many famous physicists had the great prejudice against Heisenberg's approach
to quantization. So Shr\"odinger's wave mechanics was considered by many of them 
as the end of quantum mechanics; as the possibility to describe `quantum experiments'
by using classical theory of partial differential equations (especially strong
anti Heisenberg-Bohr comments were done by Einstein and Wien).

Hence, Heisenberg and  Bohr must find some strong arguments in favour of Heisenberg
approach or disappear from quantum scene. Moreover, the whole quantum spectacle
could be ended with quite trivial final: instead great mystery --- simply a part of well 
established theory of partial differential equations. If all those circumstances
be taken into account, it would be clear how N. Bohr created complementarity
principle with all its NO-GO consequences.

So Bohr's complementarity was a kind of {\it individual complementarity.} Complementary features
were regarded to individual physical systems. It is a pity that contextualists N. Bohr and W. Heisenberg related 
the uncertainty relation not to some special class of measurement procedures of the position and momentum
described by quantum formalism, but to the position and momentum of an individual elementary particle.
This imply the prejudice that the position and momentum even in principle could not be determined
simultaneously and, moreover, that it is even in principle impossible to assign such a physical property,
e.g. position or momentum, to e.g. electron: "electron does not have trajectory."

In fact, the only possible conscious interpretation of Heisenberg's uncertainty 
principle is the statistical contextual interpretation, see e.g. [12], [13]. 
It is impossible to prepare
such an ensemble of elementary particles that dispersions of both position and momentum 
observables would be arbitrary small. Everybody would agree that only this statement
can be verified experimentally. Contextualism has to be {\it statistical contextualism}
and, consequently, complementarity has to be {\it statistical contextual complementarity.}
Such contextualism and complementarity do not contradict to the possibility of finer
description of reality than given by quantum theory.

The complex of experimental physical conditions must be split into two complexes --
a preparation procedure and a measurement procedure, see e.g. [14], [15].
A preparation procedure produces a statistical ensemble of physical system. Then
a measurement device produces results -- values of a physical observable. Nonzero 
dispersion of this random variable does not imply that individual physical system
does not have objective properties that produce values of the observable via
(\ref{H3}). Thus contextual statistical interpretation can be, in principle,
extended to {\it contextual statistical realist interpretation.}

Of course, individual complementarity can be used as an argument against the possibility 
to create finer description of physical reality than given by quantum mechanics.
But statistical complementarity cannot be used as such an argument. By contextual 
statistical interpretation it is not forbidden in principle to create such preparation and
measurement procedures that position and momentum would be measured with dispersions
$\Delta q$ and  $\Delta p$ such that $\Delta q \Delta p < \frac{h}{2}.$ Of course, such a statement
should immediately induce a storm of protests with reference to the principle of complementarity.
However, we again recall that the right complementarity principle is contextual and statistical.
It is about some class of measurement and preparation contexts described by quantum formalism.
In particular, we do not consider quantum formalism as a kind of {\it complete} physical
theory.

However, as far as we cannot perform such experiments for elementary particles, it 
is really impossible to reject Bohr's principle of individual complementarity. \footnote{Well, we may produce
various realistic models underlying quantum formalism, for example,
Bohmian mechanics, stochastic electrodynamics. However, it seems that all such models would be 
more or less automatically rejected by quantum community.}
What can we do?

We can study consequences of the general statistical contextuality and try to demonstrate that
some distinguishing features of quantum theory that are typically associated with individual
complementarity, NO-GO complementarity, are, in fact, simple consequences of statistical complementarity,
GO-DEEPER complementarity. I did this in a series of papers, see e.g.  [16], [17].
The main consequence of these investigations is that `waves of probability' can be produced in the 
general situation (including macroscopic systems) due to combination of a few preparation 
contexts.\footnote{In the two slit
experiment we consider the screen with slits as a part of an ensemble-preparation device
and the screen with photoemulsion as a measurement device.} Thus such `waves'
are not directly related to some wave features of objects participating in an experiments.
Moreover, our investigation demonstrated that in some experiments there can be created 
other types of probabilistic waves, namely hyperbolic waves of probability.

In particular, our contextual probabilistic investigations demonstrated that contextual
complementarity,  wave-particle dualism, is not rigidly coupled to microworld. Thus
we can, in principle, perform experiments with macro systems that would demonstrate
`wave-particle duality', but not of macro objects, but contexts. 

Such a 
numerical experiment,  a classical analogue of the well known  quantum two-slit
experiment, was performed in [18]. Charged particles are scattered on flat screen with two slits
and hit the second screen. We show that the probability distribution on the
second screen when both slits are open is not given by the sum of 
distributions for each slit separately, but has an extra interference term
that is given with the quantum rule of the addition of probabilistic
alternatives. We have two theoretical
descriptions of this experiment: 1) quantum-like statistical description; 2) Newtonian
classical description. Both theories give the same statistical distribution of spots
on the registration screen. Quantum-like theory operates with complex waves of probability;
there is uncertainty, Heisenberg-like, relation for position and momentum. Of course,
this relation is the statistical one. 

Suppose now that some observer could not provide
the verification of Newtonian description, e.g. such an observer is a star-size
observer and its measuring device produces nonnegligible perturbations of our macroscopic
charged balls. Such an observer might speculate on  impossibility to find objective
phase-space description and even about waves features of macroscopic balls.
This experiment might be used as (at least indirect) argument against the orthodox Copenhagen 
NO-GO in experiments. 

We recognize that at the moment there is one (and seems just one) argument supporting
individual complementarity, namely Bell's inequality.\footnote{Some people
use this framework to support quantum nonlocality.}  However, I have great doubts that the
experimental violation of Bell's inequality can be interpreted as an argument 
against the possibility of HV ontic description (or even against realism) or
locality, see e.g. my works [19],[20] (see also papers of L.Accardi, L. Ballentine,
W. De Myunck in  [21]; see also [22] on contextual statistical realist
interpretation of GHZ-paradox.

Finally, we remark that the possibility of (\ref{H3})--description implies that
`quantum randomness' does not differ essentially from `classical randomness'
Of course, this contradict to orthodox quantum views to randomness as {\it fundamental
or irreducible randomness.} Unfortunately, I could not understand the latter ideas.
Instead of fundamental irreducible quantum  randomness, I prefer to consider well understandable 
theory of context (complex of experimental physical conditions) depending probabilities.

A new fundamental principle that we propose can be called:

{\bf The principle of contextual relativity of probabilities.}

Contextuality means that all probabilities depend on complexes of 
physical conditions, ${\cal S},$
-- contexts: ${\bf P}(E) \equiv {\bf P}_{{\cal S}}(E).$ It is meaningless
to speak about probability without to determine complex of physical 
conditions. This is clear idea looks even trivial after it has been
formulated. However, we remark that the conventional probability theory
based on Kolmogorov measure theoretical axiomatics, 1933, [], is not contextual.
In Kolmogorov's theory  the probability space can be fixed once for ever. 
We need not remember that probabilities depend on complexes of physical conditons
and can use just the symbol of abstract probability ${\bf P}.$

\footnote{We remark that
A. Kolmogorov  by himself understood well contextual dependence of probabilities, see
section 2 on experimental applications of his theory in  []. However, this contextuality
was not present in his mathematical formalism. It was the terrible 
mistake of Kolmogorov. In fact, he tried to improve the situation and published
paper [] in that he noticed that probabilities are contextual probabilities and 
even used corresponding symbol. However, the paper was published only in Russian
and even in Soviet Union it was forgotten. Recently the pupil of Kolmogorov, 
A. Shiryaev, paid my attention to this paper.}

\section{Citation with comments}

In this section we shall present some citations on orthodox quantum theory and our
contextual statistical realist comments. We use, in particular, collections of Bohr's
views presented in papers of H. Folse and A. Plotnitsky, see [23], [5].

(S1)"In contrast to ordinary mechanics, the new quantum mechanics does not deal with a space-time description of the motion of
atomic particles. It operates with manifolds of quantities which replace the harmonic
oscillating components of the motion and symbolize the possibilities of transition between stationary states
in conformity with the correspondence principle", N. Bohr.

This is simply the recognition of the restrictiveness of the domain of applications of quantum theory.
I would like to interpret this as the recognition of incompleteness of quantum theory. However,
it was not so for N. Bohr:

(S2) " ... the quantum postulate implies that any observation of atomic phenomena will involve an interaction 
with the agency of observation not be neglected. Accordingly, an independent reality in the ordinary
physical sense can neither be ascribed to the phenomena nor to the agencies of observation," N. Bohr.

The first part of this citation is the manifestation of contextuality. However, I cannot
understand what kind of logic N. Bohr used to proceed to the second part. The second part
can be interpreted as the declaration of the impossibility of objective, ontic description
of reality. 

(S3) "... to reserve the word phenomenon for comprehension of effects observed under given
experimental conditions... These conditions, which include the account of the properties and manipulation
of all measuring instruments essentially concerned, constitute in fact the only basis for the definition of
the concepts by which the phenomenon is described," N. Bohr.

I would agree if the last sentence would be continued as "is described in quantum formalism."

(S4) "...by the very nature of the situation prevented from differentiating sharply between an independent 
behaviour of atomic objects and their interaction with the means of observation 
indispensable for the definition of the phenomena," N. Bohr.

I would agree if the last sentence would be continued as "of the phenomena described by quantum formalism."

(S5) "There are two forms in which quantum mechanics may be expressed,
based on Heisenberg's matrices and Schr\"odinger's wave function respectively. The second of these
is not connected directly with classical mechanics. The first is in close analogy with classical
mechanics, as it may be obtained from classical mechanics simply by making the variables of classical mechanics
into non-commuting quantities satisfying the correct commutation relations." P.Dirac, [24].

No Comment...

\section{On romantic interpretation of quantum mechanics}

Finally, we ask: "Why the realistic interpretation is not so popular in quantum community?"

The common opinion: this is the direct consequence of experiments with elementary particles.
Well, I do not think so. Of course, interference experiments with massive particles should 
induce reconsideration of methods of classical statistical mechanics. As well as discreteness of 
energy levels should induce reconsideration of classical continuous real model of physics. 
However, such reconsiderations would be merely mathematical. And it seems (at least for me)
that they were merely mathematical. Observables take only  discrete values -- consider
`discrete' number systems instead of `continuous'. The standard probability calculus (created
for one fixed sample space prepared under stable physical conditions) does not work -- create
the new one. And this was done and very successfully. However, there were no reasons to create
new quantum philosophy, based on Bohr's principle of complementarity and the individual interpretation of
Heisenberg's uncertainty relations. Nevertheless, such a new philosophy was invented (merely by N. Bohr)
and, moreover, it was recognized as philosophy of modern physics. Since this recognition, all realistic
models for experiments with elementary particles were more or less automatically rejected,
see e.g. A. Lande's statistical contextual realistic model for diffraction [25]. Lande's model looks 
quite natural; here we need not apply to wave-particle dualism, collapse and so on... It was simply
rejected. Bohmian mechanics, see e.g.  [26], -- well, it has its disadvantages, but merely, mathematical. 

Finally, Bell's inequality arguments
were interpreted as they should be interpreted in the orthodox quantum framework, despite very strong
counter-arguments. If all these counter-arguments be taken into account, Bell's inequality activity
would look very strange, as a kind of mystification.

I suspect that the main reason for this rather strange situation in modern physics is the great attraction
of romantic spirit of orthodox quantum philosophy. All these nonreal things, wave-particle dualism, collapse,
nonlocality, irreducible randomness, were  attractive for some of creators as well as further
quantum generations. These are different stories to discuss merely mathematical modification
of real-continuous model of classical statistical mechanics or to declare scientific revolution. 
So orthodox Copenhagen interpretation was a kind of {\it romantic stage} in the development of physics.
It is always not easy argue against romanticism. Probably we need not do this. I hope that realism
(as the history of literature shows us) would (sooner or later) come to physics.

I would like to thank J. Bub, P. Lahti, W. De Baere, A. Plotnitsky and I. Pitowsky for
extended discussions on the interpretation of quantum mechanics. Despite their critisism,
these discussions were very important for me; in particular, to understand better
Copenhagen Interpretation (or it would be better to say Copenhagen Interpretations) 
of quantum mechanics. Especially important role played discussions with A. Plotnitsky
including deep analysis of Bohr's and Heisenberg's views. I would like to thank 
him for his long `private lectures' on Bohr's complementarity. On the other hand,
it is important to underline that my views presented in this paper strongly differ
from his views (that are presented e.g. in [5]).

We underline that our analysis of views of N. Bohr and H. Heisenberg demonstrated that
by using the statistical interpretation of the principle of complementarity and 
uncertainty relations we escape the contradiction between the quantum formalism and
realism. Statistical approach to Bohr's contextualism does not contradict to the possibility
to construct a realist prequantum model. And recently I, finally, constructed such a model
(in spite of all {\bf no-go} theorems -- e.g., von Neumann, Kochen and Specker,...,
Bell,...), see [27]. In our model both
classical and quantum spaces are rough images of the fundamental
{\bf prespace.} Quantum mechanics cannot be reduced to classical one. But the realist
prespace model (inducing both classical and quantum representations of reality) can be constructed.
In such a model both classical and quantum
representations induce reductions of prespace information.

{\bf References}

1. H. Atmanspacher, R. C. Bishop and A. Amann, Extrinsic and intrinsic irreversibility in 
probabilistic dyanamical laws. {\it Proc. Conf. "Foundations of Probability and Physics",}
V\"axj\"o-2000, Sweden; editor A. Khrennikov, WSP, Singapore, p. 50-70, 2001.

2.  A.Yu. Khrennikov, {\it Non-Archimedean analysis: quantum
paradoxes, dynamical systems and biological models.}
Kluwer Acad. Publishers, Dordreht, 1997.

3.   V. S. Vladimirov,  I.  V. Volovich, and  E. I. Zelenov, 
{\it $p$-adic Analysis and  Mathematical Physics}, WSP, Singapore, 1994.

4.  A. Yu. Khrennikov,  {\it p-adic valued distributions and their applications to
the mathematical physics}, Kluwer Acad. Publishers, Dordreht,
1994.

5. A. Plotnitsky, Reading Bohr: Complementarity, Epistemology, Entanglement, and Decoherence.
{\it Proc. NATO Workshop "Decoherence and its Implications for Quantum Computations",}
eds. A. Gonis and P. Turchi, p.3-37, IOS Press, Amsterdam, 2001.

6. B. L. Cline, {\it The questioners: physicists and the quantum theory.}
Thomas Y. Crowell Company, New-York, 1965.

7. W. Heisenberg, {\it Physical principles of quantum theory.}
Chicago Univ. Press, 1930.

8. W. Heisenberg, {\it Zeits. f\"ur Physik,} {\bf 33},  879 (1925).

9. N. Bohr, Kramers, Slater, {\it Philosophical Magazin,} {\bf 47}, 485 (1924);
{\it Zeits. f\"ur Physik,} {\bf 24}, 69 (1924).

10. E. Schr\"odinger, {\it Ann. Physik,} {\bf 79}, 361 (1926).

11.  E. Schr\"odinger, {\it Ann. Physik,} {\bf 79}, 489 (1926).

12. H. Margenau, {\it Phil. Sci.,} {\bf 25}, 23 (1958).

13. L. E. Ballentine,  The statistical interpretation of quantum mechanics,
{\it Rev. Mod. Phys.,} {\bf 42}, 358--381 (1970). 

14. P. Busch, M. Grabowski, P. Lahti, {\it Operational Quantum Physics.}
Springer Verlag, 1995.

15. L. E. Ballentine, {\it Quantum mechanics.} Englewood Cliffs, 
New Jersey, 1989.

16. A. Yu. Khrennikov, Linear representations of probabilistic transformations 
induced by context transitions. {\it J. Phys.A: Math. Gen.,} {\bf 34}, 9965-9981 (2001).

17. A. Yu. Khrennikov, Origin of quantum probabilities. {\it Proc. Conf.
"Foundations of Probability and Physics",} V\"axj\"o-2000, Sweden; editor A. Khrennikov,  
p. 180-200, WSP, Singapore (2001).

18.  A. Yu. Khrennikov,  Ya. I. Volovich, {\it Numerical experiment on interference
for macroscopic particles.} Preprint quant-ph/0111159, 2000.

19.  A. Yu. Khrennikov, {\it Interpretations of Probability.}
VSP Int. Sc. Publishers, Utrecht/Tokyo, 1999.

20.  A. Yu. Khrennikov, A perturbation of CHSH inequality induced by fluctuations 
of ensemble distributions. {\it J. of Math. Physics}, {\bf 41}, N.9, 5934-5944, (2000).

21. {\it Proc. Conf. "Foundations of Probability and Physics",}
V\"axj\"o-2000, Sweden; editor A. Khrennikov, WSP, Singapore, 2001.

22. A. Yu. Khrennikov, Contextualist viewpoint to Greenberger-Horne-Zeilinger paradox. {\it Phys. Lett.},
A, {\bf 278}, 307-314 (2001).

23. H. Folse, {\it Bohr's best bits.} Preprint, 2001.

24.  P. Dirac, On the analogy between classical and quantum mechanics.
{\it Rev. Mod. Phys.}, {\bf 17}, 195- 199 (1945).

25. A. Lande, {\it New Foundations of Quantum Mechanics.} Cambridge Univ. Press, Cambridge, 1965.

26. D. Bohm  and B. Hiley,  {\it The undivided universe:
an ontological interpretation of quantum mechanics.} Routledge and Kegan Paul, 
London, 1993.

27. A. Yu. Khrennikov, Contextual approach to quantum mechanics and the theory of fundamental
prespace. quant-ph/0306003.

A. Yu. Khrennikov, Classical and quantum spaces and rough images of teh fundamental prespace.
quant-ph/0306069.

\end{document}